\documentstyle[epsfig]{elsart}
\begin{document}
\begin{frontmatter}
\title{Third flow component as QGP signal}
\author[Bg,Bp,Ack]{L.P. Csernai} and
\author[Bg]{D. R\"ohrich}
\address[Bg]{Department of Physics, University of Bergen\\
All\'egaten 55, 5007 Bergen, Norway}
\address[Bp]{KFKI - Research Inst. for Particle and Nuclear Physics\\
P.O.Box 49, 1525 Budapest 114, Hungary}
\thanks[Ack]{Partially supported by Research Council of Norway (NFR) and by the Alexander von Humboldt Foundation.}

\begin{abstract}
A review of earlier fluid dynamical calculations with QGP show a {\em softening}
of the directed flow while with hadronic matter this effect is absent. 
The effect shows up in the reaction plane as enhanced emission which is 
orthogonal to the directed flow. Thus, it is not shadowed by the 
deflected projectile and target. As both of these flow components are in the
reaction plane these form an enhanced 'elliptic flow' pattern.
Recent experimental data at 11 AGeV and above show the same {\em softening}, 
hinting at QGP formation.
\end{abstract}
\end{frontmatter}

\section{Introduction}

It is widely known that fluid dynamics (FD) is governed by the Equation of
State  (EoS) of the matter, and the analysis of the resulting flow
patterns turned out to be one of the best tools to extract the EoS
from the outcome of a heavy ion collision. The final event shape must 
carry the information about the pressure development during the
collision including the early stages of the collision.

The phase transition to the Quark-Gluon Plasma (QGP) is connected to a
decrease in pressure according to most theoretical estimates, not only in
strong first order phase transition models, but even if we have a smooth
but rapid gradual transition. This reduced pressure and
temperature around the phase transition threshold is known for long 
(see e.g. \cite{holme}), and it is emphasized recently as a QGP signal, 
i.e., the "soft point" of the EoS, which should be possible to observe in
excitation functions of collective flow data\cite{rpm95,shur95}.
It was pointed out earlier \cite{hpg93} that the decrease in the
out of plane (squeeze out) emission is even more sensitive to the
pressure drop and it decreases due to plasma formation stronger than the
inplane collective flow. 

Here we want to point out another consequence of the same softening in the
EoS, which is a new distinct flow pattern, which can be seen in almost all
theoretical FD calculations with QGP formation, but which was not discussed
earlier\cite{asc91,hpg93,bcl94,rpm95,bdm97}.
\bigskip

\section{Third flow component in fluid dynamics}

Up to now two basic flow patterns are predicted and detected: (i) the
directed transverse flow in the reaction plane, or side-splash, or bounce
off, which is most frequently presented on the well known $P_x$ vs. $y$
diagram and seen at all energies in heavy ion collisions from energies of 30
A.MeV to 165 A.GeV, \cite{e877,liu97,na98,wa98} 
and the (ii) squeeze-out effect which is an enhanced emission of particles
transverse to the reaction plane at center of mass (CM) rapidities. 

At lower energies the directed transverse flow resulted in a smooth,
linear $P_x$ vs. $y$ dependence at CM rapidities. This straight line
behavior connecting the maximum at $y_{proj}$ and the minimum at $y_{targ}$
was so typical that it was used to compare flow data at different beam
energies and masses.

If QGP is formed, strong and rapid equilibration and stopping takes place,
and close to one-fluid behavior is established.  Stopping is stronger than
expected, and Landau's fluid dynamical model is becoming applicable for
central collisions of massive heavy ions.  The soft and compressible QGP
forms a rather flat disk orthogonal to the beam axis which is at rest in the
CM system. Then this disk starts to expand rapidly in the direction of the
largest pressure gradient, i.e., forward and backward. Thus, the not fully
Gaussian shape of the measured rapidity spectra can be interpreted as a fluid
dynamical bounce back effect (Landau model) in contrast to the transparency
otherwise assumed in kinetic models.  Unfortunately in central collisions we
can not distinguish the two effects from one - another. Both lead to a
spectrum elongated in the beam direction.

At small but finite impact parameters, however, this disk is tilted, and the
direction of fastest expansion, will deviate from the beam axis, will stay
in the reaction plane, but point in directions opposite to the standard
directed transverse flow.  Since pressure does not play a role in
transparency, transparency cannot explain such deviation from the beam
direction! This third flow component develops purely from the large
pressure gradient at full stopping of the strongly Lorentz contracted
intermediate state. So, at the same time as the primary directed flow
is weakened by the stronger Lorentz contraction at higher energies, this third
flow component is strengthened by increased Lorentz contraction.
These two flow components together form the 'elliptic 
flow'.\cite{na98,pw98,so99}

\begin{figure}
\centerline{\epsfig{file=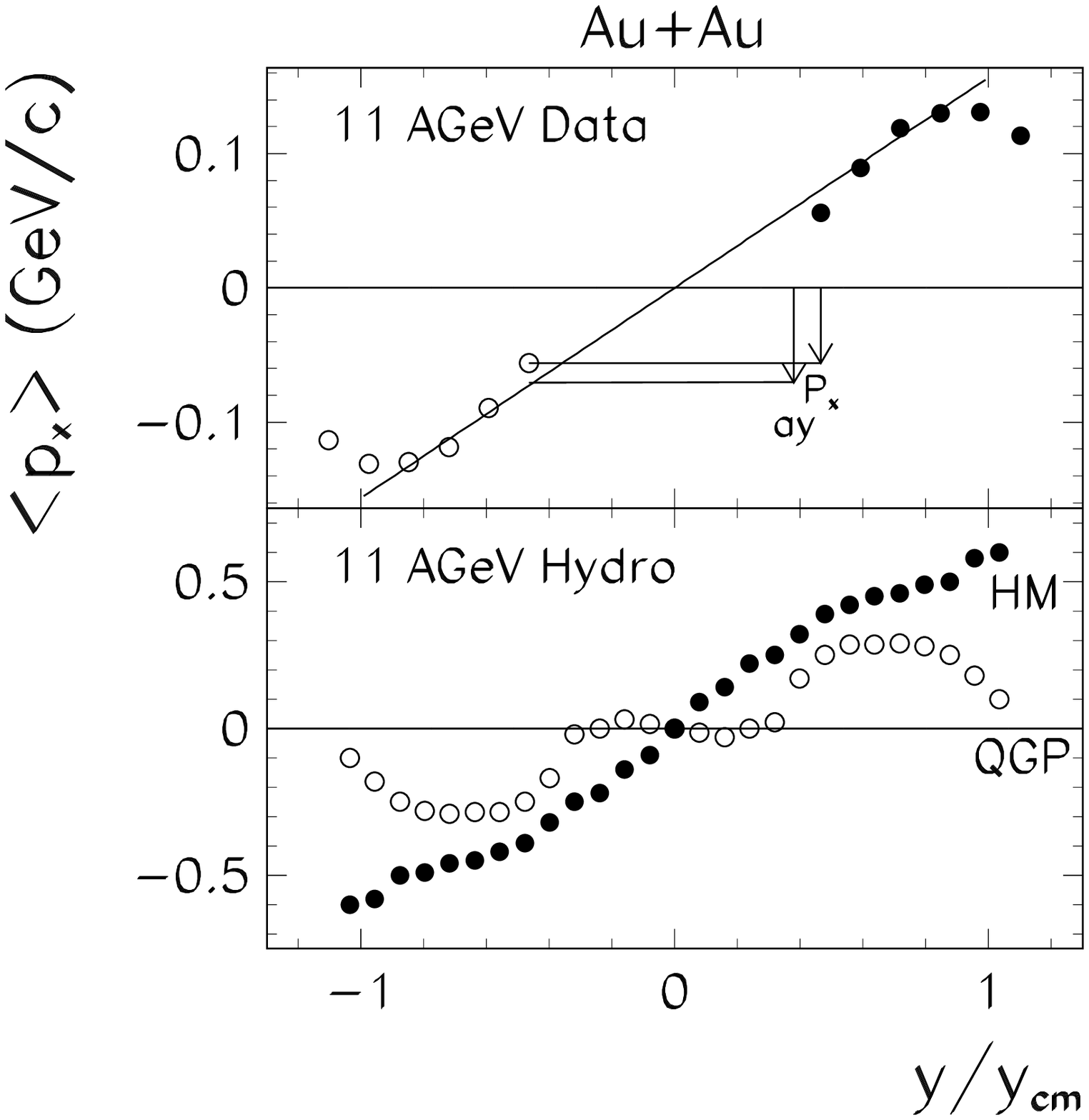,height=20.0cm,width=12cm}}
\vspace{-7cm}
\caption{
Upper part: Definition of the measure {\em softening}, $S$, describing the 
deviation of $P_x(y)$ or $v_1(y)$ from the straight line behavior,
$ay$, around CM. $S$ is defined as $|ay-P_x(y)|/|ay|$. The lower figure shows 
a typical example for fluid dynamical calculations with Hadronic and QGP EoS
\protect\cite{hpg93}. QGP leads to strong softening, $\sim 100\%$.}
\label{f1}
\end{figure}

On the $P_x$ vs. $y$ diagram \cite{do85} this component shows up as a smaller,
negative flow component at small CM rapidities. Such a third flow component
is seen clearly in Fig. 3 of \cite{hpg93} (see Fig. \ref{f1}, lower part),  
Fig. 8 of \cite{bcl94},
Fig. 6b of \cite{rpm95} and Fig. 6 of \cite{bdm97} at or slightly below 0.5
$y/y_{cm}$ if QGP formation was allowed during the calculation. In sharp
contrast, the solutions with hadronic EoS did not show this effect, and
the maximum and minimum of the $P_x$ curve could be connected with a rather
straight line. This straight line behavior is typical at all flow
results below 11 A.GeV beam energy (Fig. \ref{f2}).
In some of the FD calculations with
QGP the secondary peak at small CM rapidities is not seen, but a tendency
is obvious, and the deviation from the hadronic smooth line behavior
is apparent. This can be seen clearly in Fig. 3 of \cite{asc91}, and
Figs. 6a and 6c of \cite{rpm95}. This indicates that the strength of
this effect is also impact parameter and beam energy dependent, and the
third flow component shows a relative maximum at the same energy when
the primary directed flow is at its minimum\cite{rpm95}. Note that all
these FD calculations were done much before the experiments.  The first 
quantitative flow predictions \cite{asc91} preceded the experiments by 
as much as 6 years (!) and gave rather good agreements with the data.

To have a quantitative measure of the softening at small CM rapidities
($y_{CM}=0$) for a symmetric (A+A) collision, on the $[P_x,y]$ plane we draw
the steepest straight line, $ay$, through the CM point which is tangent to
the $P_x(y)$ curve (Fig. \ref{f1} upper part). 
The $P_x(y)$ curve and the straight line are usually
tangent to each other at $y=0$, and at finite $y$ the $P_x(y)$ curve is under
the straight line, $|ay| > |P_x(y)|$.  In case of softening at low rapidities
the straight line is tangent to the $P_x(y)$ curve at two points, $y_1$ and
$-y_1$ and at smaller rapidities, and at smaller but finite rapidities,
$0<|y|<|y_1|$, the $P_x(y)$ curve is under the straight line, $|ay| >
|P_x(y)|$. The relative deviation of these two quantities, $S(y) \equiv
|ay-P_{x}(y)| / |ay|$ has a maximum at some rapidity, 
$0<|y=y_{\mathrm{max}}| < |y_1|$,
and we will use this quantity, $S(y_{\mathrm{max}})$, 
to characterize the softening of
EoS. If $S= 100\%$ this means that $P_x(y)$ vanishes somewhere between
$0$ and $|y_1|$. If $S > 100\%$ this means that $P_x$ is inverted (shows
negative flow) at low rapidities.  
\medskip

\begin{table}[hbt]
\begin{center}
\begin{tabular}{lrr}
\hline
Publication & QGP & HM \\
\hline
\cite{rpm95} Au+Au 3.5  A.GeV $b=3$fm          &         40\%&          0\%\\
\cite{rpm95} Au+Au   5  A.GeV $b=3$fm (soft pt.)&160$\pm$30\%&  0 $\pm$20\%\\
\cite{rpm95} Au+Au 11.7 A.GeV $b=3$fm          & 70 $\pm$10\%& 30 $\pm$30\%\\
\cite{hpg93} Au+Au 11.6 A.GeV $b=5$fm          &100 $\pm$10\%&          0\%\\
\cite{bcl94} Au+Au 11.6 A.GeV $b=0-0.5b_{\mathrm{max}}$ &180 $\pm$25\%&
 30 $\pm$30\%\\
\cite{bdm97} Au+Au   11 A.GeV $b=3$fm  6 fm/c  & 70 $\pm$30\%&           - \\
\cite{bdm97} Au+Au   11 A.GeV $b=3$fm 7.2 fm/c & 50 $\pm$35\%&           - \\
\cite{asc91} Pb+Pb 160  A.GeV $b=4$fm 3.8 fm/c & 80 $\pm$10\%& 50 $\pm$45\%\\
\hline
\end{tabular}
\end{center}\bigskip
\caption[]{
Softening in fluid dynamical model calculations which included the possibility
of QGP formation in the EoS assumed. When Hadronic Matter (HM) EoS was
also presented in the same calculation the resulting softening is also shown
for this case. All HM calculations are consistent with zero softening, while
calculations with QGP show softening values from 50 to 180\%.
(In \cite{bdm97} the non-equilibrium 3-fluid hydro is applied with complex 
EoS, i.e. hadron gas EoS for the two baryon fluids (no QGP), and EoS
with QGP-hadron phase transition, which is used only for the third, baryon-free
fluid. Although the effects like directed flow are provided by baryons
the flow effect arises from the pressure of the central baryon free component.
Nevertheless, non-equilibrium effects in 3-fluid hydrodynamical models
could mimic the effect caused by the QGP in ideal 1-fluid models, and so
the comparison of this model with other fluid dynamical models is somewhat
arbitrary.)
}
\label{tab:1}
\end{table}
\bigskip

Thus in FD calculations, this effect is clearly predicted since 1991, in
all calculations, without being noticed or discussed up to now. Nevertheless,
all FD calculations are consistent in predicting an observable softening
if and only if QGP EoS is included in the calculation.
\bigskip

\section{
Deviation from straight line behavior in string models}

Some of the recent flow data are analyzed by the Fourier expansion method
\cite{o92,pw98}. Unfortunately, it is not possible to convert or compare 
these data to the earlier $P_x$ analysis, but the results indicate that
the $v_1$ vs. $y/y_{cm}$ has a straight line behavior at low rapidities
in the absence of QGP (or softening) in the EoS. Thus, just as in the
$P_x$ analysis we consider that deviation from this straight line behavior
indicates a stronger contraction and softening of the EoS.  To make
the comparison more quantitative would only be possible if data showing
such deviation from the straight line behavior would be analyzed
by both methods simultaneously. Nevertheless, heavy ion collisions
at lower energies, around 2 A.GeV show straight line behavior according
to both analyses\cite{priv}. Furthermore, small relative deviations from the
straight line behavior (or softening) are expected to be identical
in the two analyses \cite{e877} since $\langle P_x \rangle = 
\langle v_1 \rangle \cdot \langle P_t \rangle$.  

We do not intend to compare FD models and non-equilibrium models like 
string models as theoretically they are based on different fundations.
However, we review how these models describe the data.
In the last decade string models showed a surprising level of resilience
by being able to reproduce more or less all measured data.  Unlike with
FD models this always happened after the data became known.  Of course during
this time the models were changed essentially and many new ingredients were
built in. Some of these changes are in fact rather essential, and change
the physical picture described by these models drastically. One of the
most important changes is the introduction of string ropes (fused strings,
or quark clusters), and only this made it possible to describe the formation
of massive formed secondaries like strange antibaryons. Recent flow data
are also stretching the string models flexibility to their limits, while
FD models as usual overestimate collective flow data.

In Table \ref{tab:2} we present a collection of string model
results regarding the low rapidity softening. 

\medskip

\begin{table}[hbt]
\begin{center}
\begin{tabular}{lr}
\hline
Publication & S \\
\hline
\cite{APos} Venus 4.12 Pb+Pb 158 A.GeV         $v_1$         &  110$\pm$15\%\\
\cite{wa98} Venus 4.12 Pb+Pb 158 A.GeV $b=5-12$fm $P_x$      &  134$\pm$15\%\\
\cite{lpx98} RQMD  2.3 Pb+Pb    2 A.GeV $b=5-8$fm Fig. 2 $v_1$ &        40\%\\
\cite{e877}  RQMD  2.3 Au+Au 11.6 A.GeV $P_x$                 &  32$\pm$20\%\\
\cite{e877}  RQMD 1.08 Au+Au 11.6 A.GeV $P_x$                 &   0$\pm$20\%\\
\cite{wa98}  RQMD  2.3 Pb+Pb 158 A.GeV $b=8-10$fm $P_x$       &   0$\pm$25\%\\
\cite{lpx98} RQMD  2.3 Pb+Pb 158 A.GeV Fig. 1  $v_1$          &         72\%\\
\cite{lpx98} RQMD  2.3 Pb+Pb 158 A.GeV $b=5-8$fm Fig. 2 $v_1$ &         57\%\\
\cite{asc91} QGSM  Pb+Pb 158  A.GeV $b=4$fm   $P_x$           &   0$\pm$30\%\\
\cite{hpg93} QGSM  Au+Au 11.6 A.GeV $b=3$fm   $P_x$           &   0$\pm$10\%\\
\cite{b95}   QGSM  Au+Au 11.6 A.GeV $b=1,...,10$fm   $P_x$    &          0\%\\
\cite{b95}   QGSM  Au+Au 11.6 A.GeV $b=11$fm         $P_x$    &         35\%\\
\cite{b95}   QGSM  Au+Au 11.6 A.GeV $b=12$fm         $P_x$    &         50\%\\
\hline
\end{tabular}
\end{center}\bigskip
\caption[]{
Softening in string model calculations. All QGSM calculations are consistent
with zero softening, with the exception of extreme peripheral collisions,
where the effects of shadowing are observable. Both Venus 4.12 calculations
yield consistently large softening at SPS energies. RQMD results show varying
results from zero to 72\%.  }
\label{tab:2}
\end{table}
\bigskip

In an attempt to fit all data with string models in ref. \cite{lpx98} RQMD
was particularly tested to which extent it is able to reproduce NA49 flow
data.  Fig. \ref{f3} \cite{lpx98} presents the $v_1$ vs. $y$ plot for 158 A.GeV
Pb+Pb data. The proton data show a very strong deviation from the straight
line behavior (indicating strong EoS softening based on analogies with $P_x$
results). The best fits with RQMD v 2.3 also reproduce some of the deviation
from the straight line behavior, but at low rapidities the model result lags
behind the SPS data appreciably, while it yields also some softening at
2 A.GeV where experiments show no sign of such behavior.

Venus and to some extent RQMD might include some of the genuine flow effects
as the increasing Lorentz contraction, increasing pressure, and increased
stopping, but these do not quite reach the extent one can have in an FD model
with QGP.

Although, the authors of ref. \cite{lpx98} conclude that in RQMD the
deviation from the straight line behavior is caused {\em mainly} by
shadowing, it is difficult to observe any systematic viewing all published
RQMD results.  The shadowing effect is also confirmed by an earlier work
\cite{b95} which actually went more into details, and showed that for 
heavy systems like Pb+Pb the shadowing starts to cause deviations from the
straight line behavior only in peripheral reactions 
($b > 0.75 b_{\mathrm{max}}$). For
lighter systems the deviation may show up already at smaller impact
parameters also. This coincides with the result shown in Fig. 2 of
\cite{lpx98}, for Pb+Pb and S+S at 158 A.GeV.

From these string model results we can conclude that shadowing effects in
reality may also result in deviations from the straight line behavior, but in
heavy colliding systems this happens only in rather peripheral reactions,
which can be excluded experimentally. In the Pb+Pb case the shadowing effect
below the used impact parameter cut of $b<8$fm is rather weak (cf.
\cite{b95}) thus cannot be solely responsible for the deviation
from the straight line behavior. We have to add that the previously discussed
purely fluid dynamical effects also show up in string model results in a
weaker form. As this is caused by the considerable softening of the EoS due
to string ropes and other massive objects, we can actually observe some level
of convergence between FD- and string models.  Nevertheless, earlier 
the authors of string model calculations did not observe the possible 
fluid dynamical component in the explanation. In a recent version of RQMD
the measured flow pattern could be reproduced by including fluid dynamical
mechanisms and an EoS.\cite{so99}

In view of the extreme variety of string model results, we cannot reach any 
conclusion whether string models consequently yield low energy softening
or not. Furthermore, it is not clear what is the general, physical reason of
the softening in those string model calculations where such a softening is
present. 
\bigskip

\section{Experimental status}

\begin{figure}
\centerline{\epsfig{file=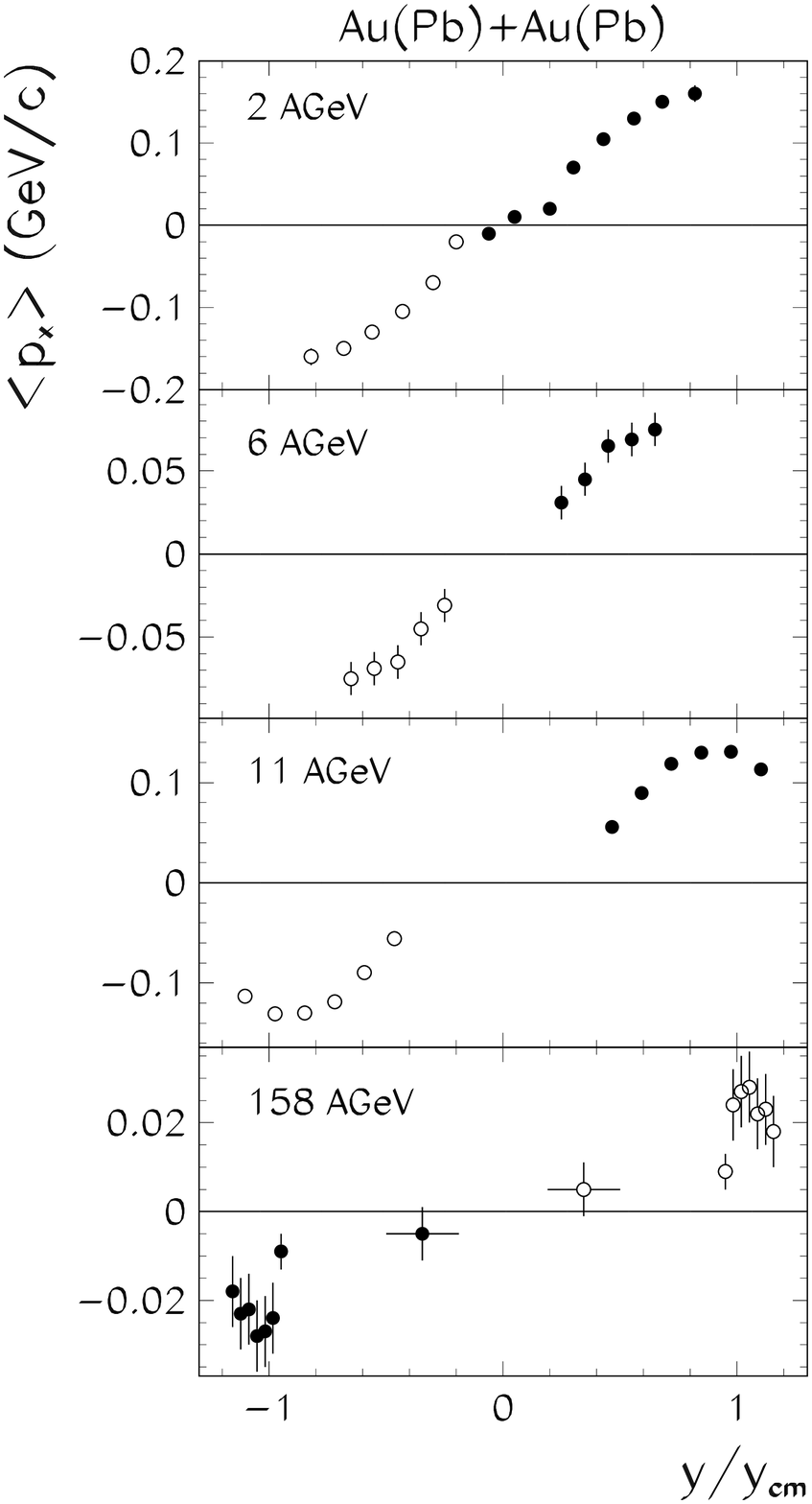,height=20.0cm,width=12cm}}
\caption{Results of transverse flow analyses, $P_x$ vs. $y/y_{cm}$,
\protect\cite{lpx98,lpx98,e877,wa98}. 158 AGeV results show strong softening.}
\label{f2}
\end{figure}

\begin{figure}
\centerline{\epsfig{file=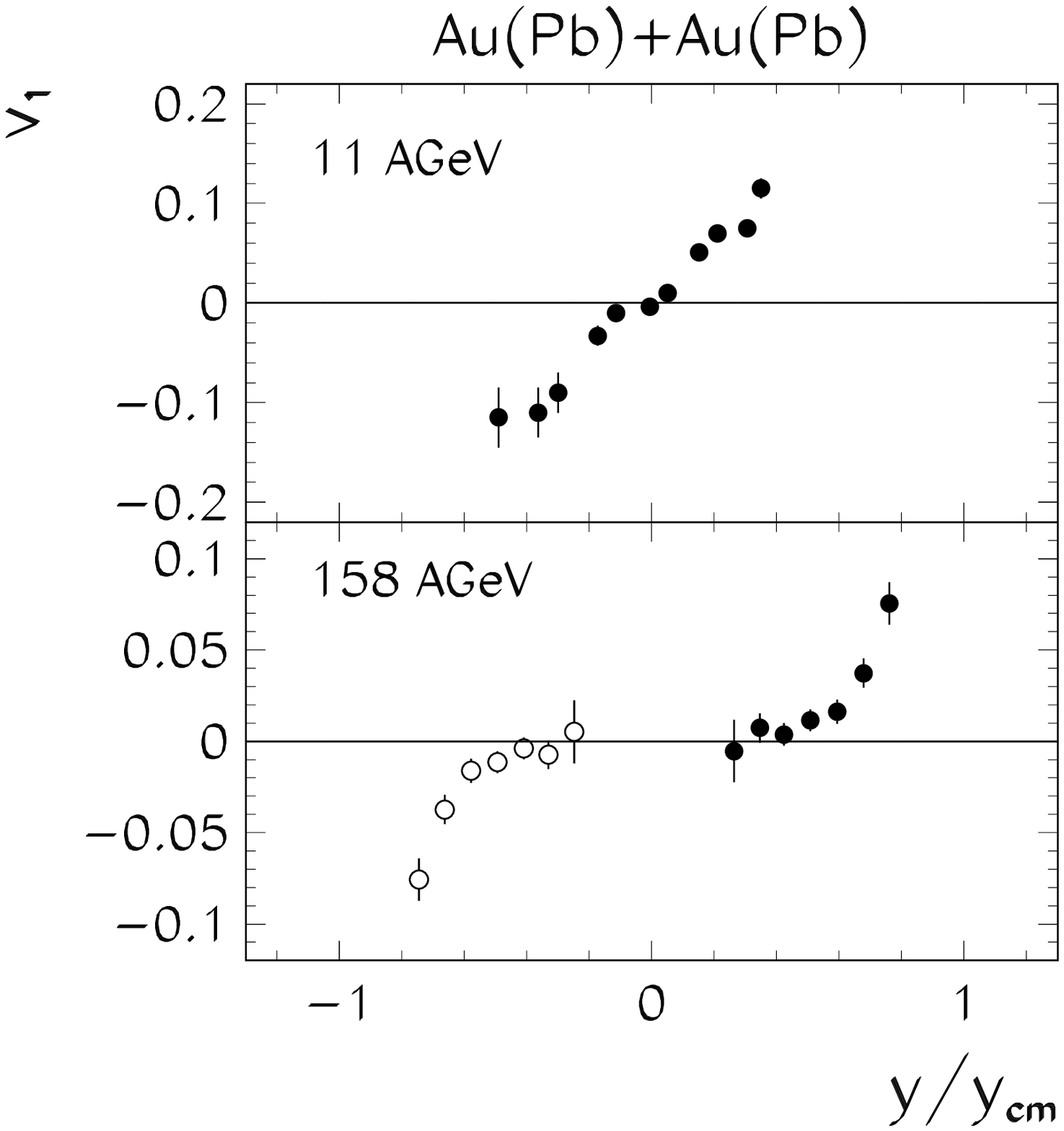,height=20.0cm,width=12cm}}
\vspace{-7cm}
\caption{Results of azimuthal Fourier analyses of directed flow,
$v_1$ vs. $y/y_{cm}$ \protect\cite{e877,na98}. Low energy (below 2 AGeV)
data show no CM softening in neither $P_x$ or $v_1$ analyses.
158 A.GeV Pb+Pb data show obvious and strong softening, $\sim 100 \%$.}
\label{f3}
\end{figure}

In connection with the string models we discussed already NA49 data.
Other experiments starting at about 2-11 A.GeV (Figs. \ref{f2} and \ref{f3})
show signs of deviation from the straight line behavior.  At 11 A.GeV 
and particularly at SPS energies the $P_x$ vs. $y$ (Fig. \ref{f2} bottom)
and the $v_1$ vs. $y$ (Fig. \ref{f3} bottom)
becomes almost flat at or close to CM rapidities. Although the inversion
of proton flow is not observed as predicted by some of the FD calculations,
the tendency from the straight line behavior is consistently seen in all
experiments (Table \ref{tab:3}).

\medskip

\begin{table}[hbt]
\begin{center}
\begin{tabular}{lr}
\hline
Publication & S \\
\hline
\cite{liu97} Au+Au 4 A.GeV $P_x$                       &     40$\pm$50\%\\
\cite{liu97} Au+Au 6 A.GeV $P_x$                       &     10$\pm$35\%\\
\cite{e877} Au+Au 11 A.GeV $P_x$  protons             &     30$\pm$20\%\\
\cite{wa98} Pb+Pb 158 A.GeV $P_x$ semicentr.          &     60$\pm$50\%\\
\cite{na98} Pb+Pb 158 A.GeV $v_1$                     &    110$\pm$60\%\\
\hline
\end{tabular}
\end{center}\bigskip
\caption[]{
Softening in directed transverse flow data. With increasing beam energy
the tendency of an increasing softening is observable. At SPS energies
the softening becomes significant. This general trend coincides with
the fluid dynamical predictions with QGP EoS. 
}
\label{tab:3}
\end{table}
\bigskip

The quantitative study of this effect in finer impact parameter resolution
may even pin down more pronounced deviations in the central rapidity
region which is important if we want to use flow data to reconstruct
the properties of the underlying EoS.
\bigskip

\section{Conclusion}

The observed flow patterns at ultra-relativistic energies indicate a new flow
pattern, which may arise from a highly Lorentz contracted and compressed
intermediate state as a consequence of an extremely soft EoS. The presently
observed deviations from the straight line behavior go beyond hadronic or
string model predictions, indicating a softer and more compressed initial or
intermediate state than hadronic models can accommodate.

A review of earlier FD calculations which included QGP shows that
such an effect was present already in all of them to a smaller or 
larger extent.  This sign of excessive softening of the EoS may indicate
that  a larger portion of the matter is transformed into a soft phase
than assumed in string models.  
The effect shows up in the reaction plane as enhanced emission which is 
orthogonal to the directed flow. Thus, it is not shadowed by the 
deflected projectile and target.\cite{hpg93,b95}
 As both of these flow components are in the
reaction plane these form an enhanced 'elliptic flow' pattern.

A review of string model results does not provide us with any clear 
conclusion about the causes or extent of the softening, although the
effect is present in some of the string model calculations.

In order to be more quantitative further and more detailed experiments are
needed, as well as further FD calculations would be beneficial.
Fluctuations, initial pre-equilibrium transparency, viscosity, etc.
may decrease the effect in FD calculations also, thus approaching
the real reaction mechanism better than recent calculations.

\newpage

\newpage

\newpage

\end{document}